\documentclass[aps,prl,twocolumn,superscriptaddress,floatfix]{revtex4-1}

\usepackage{graphicx}
\usepackage{amsmath}
\usepackage{bbold}
\usepackage[T1]{fontenc}
\usepackage{xcolor}
\usepackage{hyperref} 

\newcommand{\beq}{\begin{equation}}
\newcommand{\eeq}{\end{equation}}
\newcommand{\beqs}{\begin{equation*}}
\newcommand{\eeqs}{\end{equation*}}
\newcommand{\beqa}{\begin{eqnarray}}
\newcommand{\eeqa}{\end{eqnarray}}
\newcommand{\beqas}{\begin{eqnarray*}}
\newcommand{\eeqas}{\end{eqnarray*}}
\def\bals#1\eals{\begin{align*}#1\end{align*}}
\def\bal#1\eal{\begin{align}#1\end{align}}

\newcommand{\bcent}{\begin{center}}
\newcommand{\ecent}{\end{center}}

\newcommand{\bitem}{\begin{itemize}}
\newcommand{\eitem}{\end{itemize}}

\makeatletter
\newcommand*\bt{\mathpalette\bt@{.7}}
\newcommand*\bt@[2]{\mathbin{\vcenter{\hbox{\scalebox{#2}{$\m@th#1\bullet$}}}}}
\makeatother

\makeatletter
\newcommand*\ct{\mathpalette\ct@{.7}}
\newcommand*\ct@[2]{\mathbin{\vcenter{\hbox{\scalebox{#2}{$\m@th#1\circ$}}}}}
\makeatother

\begin{document}



\title{Anomalies in the electronic stopping of slow antiprotons in LiF}

\author{Guerda Massillon-JL}
\affiliation{Instituto de F\'{\i}sica, Universidad Nacional 
             Aut\'onoma de Mexico, 04510 Mexico City, Mexico}
\author{Alfredo A. Correa}
\affiliation{Quantum Simulations Group, Lawrence Livermore 
             National Laboratory, Livermore, 
             California 94550, USA}
\author{Xavier Andrade}
\affiliation{Quantum Simulations Group, Lawrence Livermore 
             National Laboratory, Livermore, 
             California 94550, USA}             
\author{Emilio Artacho}
\affiliation{CIC Nanogune and DIPC, Tolosa Hiribidea 76, 
              20018 San Sebastian, Spain}
\affiliation{Ikerbasque, Basque Foundation for Science, 
              48011 Bilbao, Spain}             
\affiliation{Theory of Condensed Matter,
             Cavendish Laboratory, University of Cambridge, 
              J. J. Thomson Ave, 
             Cambridge CB3 0HE, United Kingdom}

\date{\today}

\begin{abstract}
  We present first-principles theoretical calculations for the 
electronic stopping power (SP) of both protons and anti-protons in LiF.
Our results show 
the presence of the Barkas effect: 
a higher stopping for positively charged particles than 
their negatively charged antiparticles.
  In contrast, a previous study has predicted 
an anti-Barkas effect 
(higher stopping for negative charges) at low velocity 
[Qi, Bruneval and Maliyov, Phys. Rev. Lett. {\bf 128}, 043401 (2022)]. 
  We explain this discrepancy by showing that this anti-Barkas effect 
appears for highly symmetric trajectories and disappears when 
considering trajectories that better reproduce the experimental setup.
Our low-velocity results show that the SP of both protons and 
anti-proton vanish for velocities under 0.1 a.u. .
\end{abstract}


\maketitle



  Since the discovery of its thermoluminescent properties at the 
beginning of the 1950s~\cite{Daniels1953}, lithium fluoride (LiF),
in addition to being a prototypical large band-gap insulator, it
has been considered a crystal both of fundamental interest to 
understand the interaction of ionizing radiation with matter 
and of practical relevance as a dosimeter measuring 
absorbed dose in medical, space and occupational personal 
dosimetry and in environmental monitoring~\cite{Massillon2007}. 
  Studying the interaction of charged particles, like electrons, 
protons, and heavier ions, with matter is important in many 
areas of science and technology, including radiotherapy and 
nuclear and space activities. 

  Electronic stopping power (SP) is a quantity that defines the 
capability of a material to cause a charged particle to lose kinetic 
energy through electronic processes.
  SP is measured in energy per unit length traveled into matter.
  Several historic approximations, such as Fermi-Teller~\cite{Fermi1947} 
for low-velocities, Bethe~\cite{Fano1963} at large velocities and 
linear response coincide in predicting a SP proportional to the 
square of the projectile charge (\(Z^2\)). 
  Under these analytic descriptions, the SP for a particle and its 
oppositely charged antiparticle should be the same.
  Nevertheless, Barkas and co-workers reported that for a similar initial 
velocity, the range of negative pions was longer than that of positive 
pions~\cite{Smith1953,Barkas1956}.
  Such results suggested that the stopping power for the positive projectile 
is larger than that for the negative counterpart~\cite{Smith1953}: 
the so-called ``Barkas effect''.
They attributed this effect to the asymmetric polarization of the 
target electrons with respect to attraction and repulsion. 
The SP for protons and antiprotons in Si has been measured in the velocity 
range between \(4.64~\mathrm{a.u.}\) and \(11~\mathrm{a.u.}\) finding a 
contribution of order \(Z^3\) to the stopping power and the antiproton 
stopping power smaller than the proton by 19\% at low-velocities and 
3\% at high-velocities, which confirmed the 
Barkas effect in solids~\cite{Andersen1989}.


\begin{figure}
   \centering
   \includegraphics[width=0.9\linewidth]{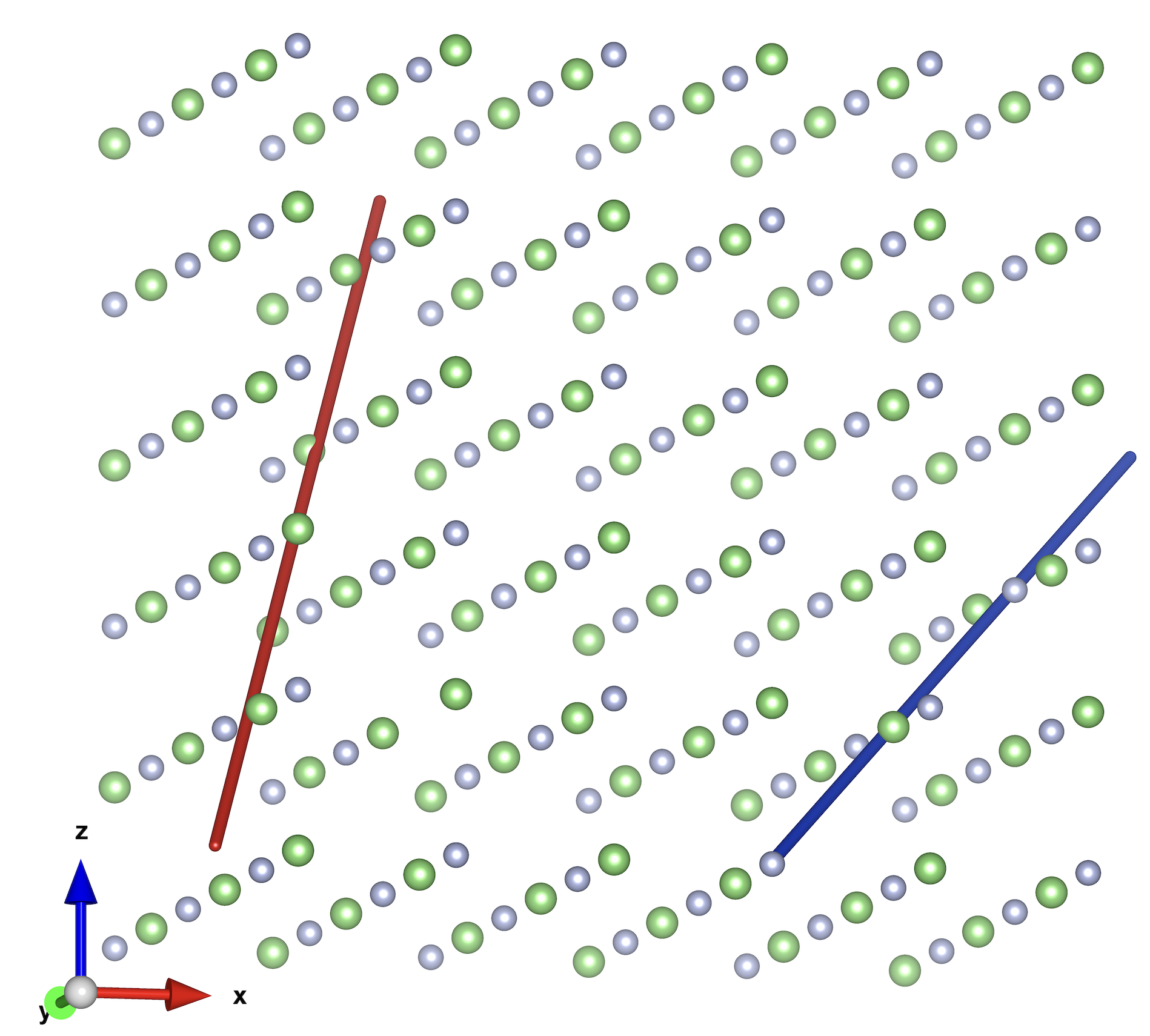}
\caption{LiF supercell of 216 atoms with two 
projectile trajectories, in a random incommensurate direction (red), 
and in a \(\langle 111\rangle\) channeling (blue), for a small 
impact parameter \(p=0.225 \text{\AA}\).
The small impact parameter produces frequent repeating close 
approaches to host atoms. (The trajectories are 3-dimensional.)
The gray and green balls represent F and Li, respectively.}
\label{fig:supercell}
\end{figure}


  Several experimental studies about SP of protons (H$^+$) in LiF 
have been reported~\cite{Bader1956,Moller2004,Bauer2005,Bauer2009}, 
but to the best of our knowledge, only one for antiprotons (H$^-$) 
by M{\o}ller et al.~\cite{Moller2004}.
  They experimentally investigated the SP for protons and 
antiprotons in LiF at velocities down to \(0.3~\mathrm{a.u.}\) 
and \(0.4~\mathrm{a.u.}\), respectively~\cite{Moller2004}. 
  They reported differences in the SP for protons and 
antiprotons of around 50\%-60\%. 
  They also concluded that the SP for the antiproton 
is almost linear with the velocity~\cite{Moller2004}. 
  Measurements of the SP of low-velocity protons 
(\(v<0.5~\mathrm{a.u.}\)) in LiF indicated that the SP vanishes 
below a certain velocity threshold of approximately 
0.1 a.u~\cite{Bauer2005,Bauer2009}.
Theoretical studies based on real-time time-dependent 
density functional theory (rt-TDDFT) simulations have 
confirmed these results \cite{Pruneda2007,Zeb2013,Qi2022}.
From theory, a strict threshold is not expected but rather
a smooth but strongly depressed SP at low velocities
\cite{Artacho2007,Forcellini2020} based on conservation
arguments~\cite{Ullah2015}.
However, for low-velocity antiprotons where only 
rt-TDDFT simulated data exist, the results are still controversial. 
For antiprotons moving in \(\langle 100\rangle\) channeling 
trajectories, a similar velocity threshold and the Barkas effect 
have been obtained from rt-TDDFT simulations using atomic 
orbitals \cite{Pruneda2007}. 
  In contrast, results from rt-TDDFT using Gaussian 
basis sets with the proton moving along a \(\langle 111\rangle\) 
channeling trajectory suggested a negative Barkas 
effect (SP for antiprotons greater than that for 
protons) at velocities below 0.25 a.u. and, remarkably, 
no velocity threshold for antiprotons \cite{Qi2022}.

  This letter presents SP results for protons and 
antiprotons in LiF via rt-TDDFT simulations within 
a pseudopotential plane-wave framework. 
  Good agreement with available experimental data is obtained 
for both protons and antiprotons. 
  A similar threshold effect is observed for both, with 
no trace of the reported anti-Barkas effect in the 
low-velocity limit, in apparent contradiction with the 
Gaussian basis simulation results.


  Simulations via rt-TDDFT were performed through the 
open-source code Qball~\cite{QBALL,Draeger2017} which is based 
on the plane-wave approach.
  In that code, the temporal evolution of the system's 
total energy caused by the dynamic of the charged particles 
is described by the time-dependent Kohn-Sham equations.
  Optimized norm-conserving Vanderbilt (ONCV) 
pseudopotentials~\cite{Hamann2013} were used to describe
the interaction between valence electrons
and core ions (Li and F ions with 3 and 7 valence 
electrons, respectively). 
  For the proton, a local Coulomb pseudopotential for
hydrogen was used, with the equivalent repulsive 
pseudopotential for the antiproton.
  The Perdew-Burke-Ernzerhof (PBE) functional~\cite{Perdew1996} 
was used for the exchange-correlation (XC) functional. It is well known 
that PBE underestimates 
the band gap, which could affect proportionally the predicted 
velocity threshold.
  However, as we will show in this letter, it does not do it 
noticeably, since what we obtain is similar to experiments on 
the scale we are studying.
  The electronic wave function evolved through the enforced 
time reversal symmetry propagator~\cite{Castro2004}. 
  The plane-wave basis set was the one corresponding to a 50 
Hartree energy cutoff. 

\begin{figure}[t] 
\includegraphics[width=0.49\textwidth]{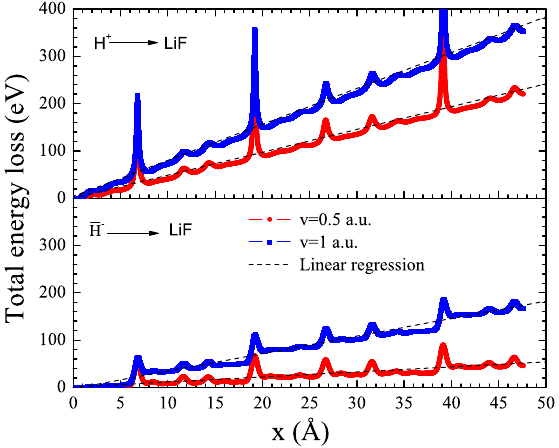}
\caption{Electronic total energy gain (projectile´s 
kinetic energy loss) versus projectile displacement $x$ 
along an incommensurate trajectory, for proton/antiproton 
(upper/lower panel), and for $v=0.5$ a.u./1.0 a.u. 
(red/blue color).}
\label{fig:EvsX}
\end{figure}

  A time step, $\Delta t = 1$ attosecond was used for 
velocities < 0.5 a.u. and for greater velocities, a 
constant displacement of $\Delta x = 0.005$ \AA\ was set 
in each integration step ($\Delta t = \Delta x/v$). 
  Both projectiles were moved in a direction incommensurate 
with the crystalline lattice~\cite{Correa2018} at constant velocity 
through a $3\times 3\times 3$ supercell of a pristine LiF 
containing 216 atoms (Fig.1~\ref{fig:supercell}). 
  The geometry of the pristine supercell was previously 
optimized at the hybrid PBE0 level rendering a lattice 
constant of 4.028 \AA\ \cite{Massillon2018}. 
  For the calculation of electronic stopping
power, the host atoms were frozen during the simulation, thereby
avoiding nuclear stopping effects by construction.
The SP \(S_\text{e} = dE/dx\) in this work was obtained as 
the slope from a linear regression of the total electronic 
energy versus the projectile displacement $x$ 
(see Fig.~\ref{fig:EvsX}).
Among different trajectory sampling techniques 
discussed in the literature for SP averaging 
\cite{Correa2018,Rafi2018,BinGu2020,Ward2024} the 
incommensurate-trajectory method has been shown to be 
the simplest and satisfactorily accurate.
Atomic units are used for \(v\) and \(\mathrm{eV/\AA}\) 
for \(S_\text{e}\) as customary in this field (1 Hatree/Bohr = 
51.422 eV/\AA).


\begin{figure}[t] 
\includegraphics[width=0.48\textwidth]{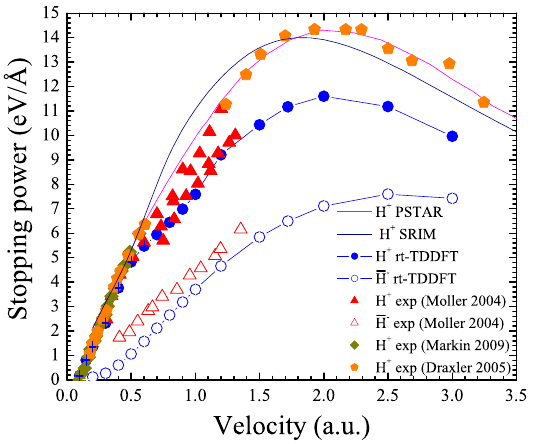}
\includegraphics[width=0.48\textwidth]{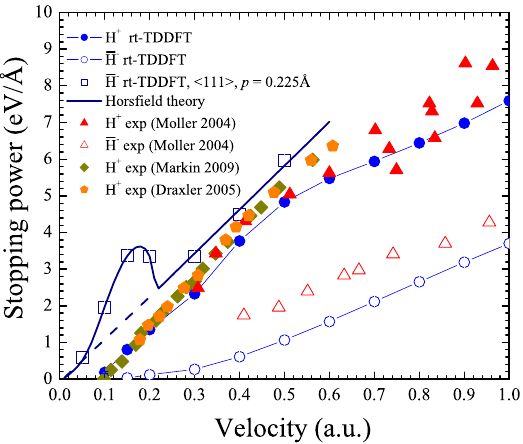}
\caption{Electronic stopping power for protons (filled symbols) 
and antiprotons (open symbols) travelling through LiF 
versus projectile velocity.
  Upper panel: Results obtained via rt-TDDFT simulations 
are compared with reported experimental results and data 
from SRIM and PSTAR.
  Lower panel: Simulation results for low-velocity proton and 
antiproton averaged trajectories are compared with a 
\(\langle 111\rangle\) channelling  trajectory for antiprotons 
with an impact parameter (closest approach to fluorine 
atom) of \(p = 0.225 \text{\AA}\).
The dark blue continuous line shows the Horsfield 
model, with contributions from the \(S_\text{d}\) formula in the 
text, which generates the shoulder at around \(1.5~\mathrm{a.u.}\), 
plus from an adjusted linear contribution from interband stopping 
(dashed extrapolated line).}
\label{fig:stopping}
\end{figure}

  Figure~\ref{fig:stopping} displays our rt-TDDFT results 
of the SP for protons and antiprotons in LiF as a function 
of the projectile velocity. 
  Note that no negative Barkas effect is observed in the 
SP for antiprotons, as the antiproton stopping is consistently 
lower than proton stopping. 
  Compared with the experiments~\cite{Moller2004,Bauer2005,Bauer2009}, 
we find good agreement with the simulation for the proton where 
a velocity threshold of $v \sim 0.1$ a.u is also obtained. 
  Relative to the SRIM~\cite{SRIM,Ziegler2010} and 
PSTAR~\cite{PSTAR,Berger1999} data, the maximum 
SP obtained in the simulation is smaller by about 2 eV/\AA. 
  This discrepancy is due to the \(1s\) electrons
of F being frozen into its pseudopotential \cite{Rafi2018,Brune2016}.
Concerning the antiproton, note a slight difference (or shift) 
with the experiment which is relatively more pronounced 
at lower velocities. 
  This discrepancy can presumably be associated with 
the uncertainties in the thickness of the foils used in the 
experiment~\cite{Moller2004}, which could result in a 
systematic error.

  Furthermore, similarly to the protons, a velocity threshold, 
where the SP vanishes, exits for the antiprotons. 
  For the antiprotons, a velocity threshold at around 
$v = 0.2$ a.u. is obtained. 
  Such a result contrasts with the hypothesis that the 
velocity threshold where the SP vanishes for the antiprotons 
should be smaller than for protons~\cite{Solleder2009}. 

  To understand the negative Barkas effect reported by 
Qi et al.~\cite{Qi2022}, we performed simulations for antiprotons 
moving at several constant velocities along the $\langle 111\rangle$ 
direction in channeling trajectories with a low impact parameter
$p=0.225$ \AA\ relative to the fluorine or lithium sites.
  This result is shown in Fig.~\ref{fig:stopping} and, interestingly, 
the negative Barkas effect is observed in the velocity range studied 
and a local maximum at velocity  below \(v = 0.3~\mathrm{a.u.}\), 
as also reported by Qi et al. for the smallest impact parameter 
$p = 0.225$~\AA\ (see inset of Fig.~2 from 
Ref.~\cite{Qi2022}). 


  According to the Fermi-Teller theory regarding capture of 
mesons~\cite{Fermi1947}, the presence of a low negative charge 
moving within one lattice parameter from an atom in an insulator 
can generate a gap state, which has been confirmed by 
calculations of Solleder et al. who showed a 
localised state in the LiF gap caused by the repulsive potential of 
the antiproton~\cite{Solleder2009}.
 They did not find an in-gap state for the proton.
  Based on the Fermi-Teller approach, at low velocity, 
there would exist a critical distance (Fermi-Teller radius, 
\(r_\text{FT} = 0.639/Z a_\text{B}\)) where the antiproton will 
be captured by the fluorine, provoking a resonance that 
locally enhances the energy loss by the antiproton~\cite{Fermi1947}.
  This extra energy loss may explain the anti-Barkas effect 
observed on the SP calculated in channeling trajectories 
reported by Qi et al.~\cite{Qi2022} at low velocity and 
low impact parameter.
  An oscillatory gap state characteristic of an antiproton traveling 
at repeated close approaches would lower the effective large gap of 
LiF, which is otherwise responsible for the threshold behavior.
  Since this state sits at the Fermi level, it would be able 
to carry electrons from the valence to the conduction band in 
analogy to the effect termed `electron elevator' reported in 
Ref.~\cite{Lim2016} and more generally modeled in 
Ref.~\cite{Horsfield2016}, providing a mechanism for 
sub-threshold behavior observed for the channeling trajectories 
assumed by Qi et al.

  To model the subthreshold structure for low-impact parameters 
and at low velocity (\(v < 0.2~\mathrm{a.u.}\), we use the theory 
developed by Horsfield et al.~\cite{Horsfield2016}, which was 
initially designed to capture the effect of a gap level formed 
by a slow projectile on the electronic stopping in insulators. 
The mentioned gap level in LiF observed 
for the antiproton (and not the proton) \cite{Solleder2009} makes 
the Horsfield model suitable to model the antiproton electronic SP. 
  This perturbative model establishes that a general gap level 
(carried by the slow projectile) generates an additive stopping 
\(S_\text{d}\) on top of the normal interband (valence-conduction) 
stopping power.
  \(S_\text{d}\) is a function of (i) the energy distance 
\(|\bar\varepsilon_\text{d}|\) below/above the unoccupied/occupied band 
with a electronic density of states \(\mathcal{D}\), 
(ii) the amplitude of this oscillation \(\eta\), and 
(iii) the passing frequency \(\omega\).
\begin{equation}
S_\text{d} \propto \sum_\ell \ell |I_\ell(\eta/\hbar 
\omega)|^2 \mathcal{D}(\bar\varepsilon_\text{d} + \ell \hbar\omega),
\end{equation}
where \(I_\ell(x) = \frac{1}{2\pi} \int_0^\pi 
\frac{\sin(\ell u + x \sin u)}{\ell + x \cos u} \sin u \,\mathrm{d}u\) 
which we denominate the Horsfield function.
  Note that \(\omega\), inverse of a passing time, is proportional 
to the velocity \(\omega = 2 \pi v / \lambda\).
  A single (conduction) parabolic band of width \(\omega\) is assumed, 
\(\mathcal{D(\varepsilon)} \propto \sqrt{\varepsilon}\) with a soft 
energy cutoff (band width) \(\Omega\).
We fix \(\eta\) and \(\bar\varepsilon_\text{d}\) at 
\(12~\mathrm{eV}\) 
and \(-6~\mathrm{eV}\), 
which respectively correspond to the amplitude and average 
relative to the bottom of the conduction band, based on Fig.~3a of 
Ref.~\cite{Solleder2009} for the antiproton gap level.
A non-linear least-square fitting results in 
\(\lambda = 0.15~\text{\AA}\), \(\Omega = 3.84~\mathrm{eV}\). 
  (For these parameters, \(\ell \geq 2\) contributions are 
vanishingly small.)
  The total electronic stopping power from this simple model 
(\(S_\text{d}\) plus an adjusted linear interband stopping) is 
depicted in Fig.~3b.

  Subthreshold behavior for the antiproton is only observed for these 
unique trajectories (channeling and small impact parameter), and our 
results for the average stopping power do not reveal subthreshold 
features.
  In this regard, we attribute the discrepancy with Ref.~\cite{Qi2022} 
to a different way of sampling trajectories.
  (The plane wave methods are more amenable to simulations at random 
directions thanks to periodic boundary conditions.) 
  This illustrates that the random direction (or incommensurate) 
averaging method is not necessarily equivalent to sampling impact 
parameters for strict channeling trajectories over a crossing plane.
  We interpreted that such theoretically described subthreshold behavior 
would not be directly observed in simple experiments since projectile 
trajectories with small impact parameters (in a channel) would be 
unfavored by nuclei scattering.


  In conclusion, by comparing systematic simulations of the 
electronic stopping power of protons and antiprotons, we find 
a positive Barkas effect; that is, the positively charged 
projectile has larger electronic stopping power than the 
negatively charged counterpart across the investigated velocity
range. 
  Both projectiles have a similar threshold velocity within 
the simulation’s resolution (at $v = 0.1$ a.u.). 
  The proton stopping is five or more times larger than the 
antiproton stopping at low velocity. 
  The relative difference reduces to less than half near the 
maxima (at $v = 2.0$ a.u. and 2.5 a.u., respectively).
  At higher velocities (>3.5 a.u.), we expect the stopping 
values of protons and antiprotons to eventually converge when 
linear theory and the Bethe limit become more exact. 
  Accurate reporting at such high velocities would require 
simulations with more explicit valence electrons within the 
pseudopotential method. 
  The antiproton results are systematically lower than 
M{\o}ller’s experiment by $\sim 1$ eV/\AA. 
  We explain the apparent discrepancy between the theoretical 
results of the previous first principles by combining 
arguments regarding the trajectory sampling method and a 
physical effect characteristic of small impact parameter 
channeling trajectories.


{\it Acknowledgments.} 
The authors thank Irving Carlos Álvarez Castillo from DGTIC-UNAM for technical support.  We acknowledge computational support from Atlas and 
Hyperion DIPC, Spain and Miztli UNAM Mexico through grant 
ref. LANCAD-UNAM-DGTIC-334. 
  This project is partially supported by the European 
Commission Horizon MSCA-SE Project MAMBA (Grant No. 101131245).
  GMJL's sabbatical leave at CIC-Nanogune in Spain was
funded by DGAPA UNAM. 
  AAC and XA work was supported by the Center for 
Non-Perturbative Studies of Functional Materials Under 
Non-Equilibrium Conditions (NPNEQ) funded by the Computational 
Materials Sciences Program of the US Department of Energy, 
Office of Science, Basic Energy Sciences, Materials Sciences 
and Engineering Division and performed under the auspices of 
the US Department of Energy by Lawrence Livermore National 
Laboratory under contract DE-AC52-07NA27344. 
EA acknowledges funding from the Spanish MCIN/AEI/10.13039/501100011033 
through grants PID2019-107338RB-C61 and PID2022-139776NB-C65, as well 
as a Mar\'{\i}a de Maeztu award to Nanogune, Grant CEX2020-001038-M. 

%

\end{document}